\documentclass[%
reprint,
linenumbers,
amsmath,amssymb,
prx,
floatfix,
]{revtex4-2}

\usepackage{graphicx}
\usepackage{dcolumn}
\usepackage{xcolor}
\usepackage{bm}
\usepackage{mathtools} 
\usepackage{physics} 
\usepackage{hyperref}
\usepackage{cleveref} 
\usepackage{changes}

\allowdisplaybreaks 

\DeclareMathOperator{\Circ}{Circ}

\begin{document}
\nolinenumbers 
\preprint{APS/123-QED}

\title{Quantum limits for precisely estimating the orientation and wobble of dipole emitters}

\author{Oumeng Zhang}
\author{Matthew D. Lew}
\email{mdlew@wustl.edu}
\affiliation{
Department of Electrical and Systems Engineering, Washington University in St.\ Louis, Missouri 63130, USA
}

\date{\today}

\begin{abstract}
Precisely measuring molecular orientation is key to understanding how molecules organize and interact in soft matter, but the maximum theoretical limit of measurement precision has yet to be quantified. We use quantum estimation theory and Fisher information (QFI) to derive a fundamental bound on the precision of estimating the orientations of rotationally fixed molecules. While direct imaging of the microscope pupil achieves the quantum bound, it is not compatible with widefield imaging, so we propose an interferometric imaging system that also achieves QFI-limited measurement precision. Extending our analysis to rotationally diffusing molecules, we derive conditions that enable a subset of second-order dipole orientation moments to be measured with quantum-limited precision. Interestingly, we find that no existing techniques can measure all second moments simultaneously with QFI-limited precision; there exists a fundamental trade-off between precisely measuring the mean orientation of a molecule versus its wobble. This theoretical analysis provides crucial insight for optimizing the design of orientation-sensitive imaging systems.
\end{abstract}

\maketitle


Since the first observation of single molecules \cite{moerner1989}, scientists and engineers have worked tirelessly to quantify precisely their positions \cite{VonDiezmann2017,Cnossen2020,Gwosch2020} and orientations \cite{Backlund2014,PhysRevLett.115.173002,lippert2017angular,zhang2018imaging,zhang2019fundamental} to probe dynamic processes within soft matter at the nanoscale. Two fundamental challenges confront these experiments: the optical diffraction limit, i.e., the finite numerical aperture of the imaging system, and Poisson shot noise associated with photon counting. In recent decades, microscopists have developed numerous technologies \cite{foreman2008determination,Sikorski:08,backer2014extending,hashimoto2015orientation,curcio2019birefringent} to measure the orientations of single-molecule (SM) dipole moments. Classical estimation theory, i.e., Fisher information (FI) and the associated Cram\'er-Rao bound (CRB) \cite{moon2000mathematical}, allows us to calculate conveniently the best-possible precision of unbiased measurements of a few parameters. However, calculating the CRB requires us to assume a comprehensive set of priors about the object and the imaging system, such as the number of sources, their positions and orientations, their emission spectra and anisotropies, an exact model of the imaging system and its detector, etc. The performances of several orientation-sensing methods have been compared using CRB \cite{foreman2011fundamental,agrawal2012limits}, but the fundamental limit of measurement sensitivity remains unexplored.

Recently, quantum estimation theory has ignited a series of studies that explore the fundamental limits of estimating the 2D \cite{tsang2015quantum} and 3D \cite{backlund2018fundamental} positions of isolated optical point sources, as well as the limits of resolving two or more sources that are separated by distances smaller than the Abb\'e diffraction limit \cite{tsang2016quantum,PhysRevLett.117.190802,Rehacek:17,ang2017quantum,tsang2019,prasad2019}. Since quantum noise manifests itself as shot noise in incoherent optical imaging systems, the quantum Cram\'er-Rao bound (QCRB) sets a fundamental limit on the best-possible variance of measuring any parameter of interest. Further, this approach provides insight into how one may design an instrument to saturate the quantum bound, thereby achieving a truly optimal imaging system \cite{tsang2016quantum,backlund2018fundamental}. However, to our knowledge, no studies exist to quantify the limits of measuring the orientation and rotational ``wobble'' of dipole emitters, which has numerous applications in biology and materials science \cite{cruz2016quantitative,backer2016enhanced,lippert2017angular,backer2019single,Ding2020}.

Here, we apply quantum estimation theory to derive the best-possible precision of estimating the orientations of rotationally fixed fluorescent molecules, regardless of instrument or technique. We compare multiple existing methods to this bound and present an interferometric microscope design that achieves quantum-limited precision. Extending our analysis to rotationally diffusing molecules, we derive bounds on estimating the temporal average of second-order orientational moments and show sufficient conditions for reaching quantum-limited measurement precision. Interestingly, {while the position and orientation of a non-moving and non-rotating dipole can be measured simultaneously with quantum-limited precision,} we find that it is impossible to achieve QCRB-limited precision when estimating {both} the {average} orientation and wobble of a molecule.

\section{Imaging Model and Quantum Fisher Information}
We model a fluorescent molecule as an oscillating electric dipole \cite{novotny2012principles} with an orientation unit vector $\bm{\mu}=[\mu_x,\mu_y,\mu_z]^\dagger=[\sin\theta\cos\phi,\sin\theta\sin\phi,\cos\theta]^\dagger$. For any unbiased estimator, the covariance matrix $\bm{V}$ of estimating the molecular orientation $\bm{\mu}$ is bounded by the classical and quantum CRB \cite{helstrom1976quantum,moon2000mathematical,ang2017quantum}
\begin{equation}
    \bm{V}_{\bm{\mu}}\succeq\bm{\mathcal{J}}^{-1}\succeq\bm{\mathcal{K}}^{-1},
\end{equation}
where $\bm{\mathcal{J}}$ and $\bm{\mathcal{K}}$ represent the classical and quantum Fisher information matrices (FI and QFI), respectively, and $\succeq$ denotes a generalized inequality such that $(\bm{V}_{\bm{\mu}}-\bm{\mathcal{J}}^{-1})$ and $(\bm{\mathcal{J}}^{-1}-\bm{\mathcal{K}}^{-1})$ are positive semidefinite. Here, we consider the orientational parameters $[\mu_x,\mu_y]$ in Cartesian coordinates. Other representations of $\bm{\mu}$ can be analyzed similarly.

If the photons detected at position $[u,v]$ follow a Poisson distribution with expected value $I(u,v;\bm{\mu})$, the entries of the classical Fisher information matrix $\bm{\mathcal{J}}$ are given by 
\begin{equation}
\label{eqn:classicalFI}
    \mathcal{J}_{ij}=\iint \frac{[\pdv*{I(u,v;\bm{\mu})}{\mu_i}][\pdv*{I(u,v;\bm{\mu})}{\mu_j}]}{I(u,v;\bm{\mu})}\,du\,dv,
\end{equation}
Note that $I(u,v;\bm{\mu})$ is a property of the imaging system, i.e., any modulation of the collected emission light generally alters the classical FI matrix. 

A fundamental bound on estimation precision is given by the quantum FI matrix, which is only affected by how photons are collected by the imaging system, i.e., its objective lens(es). For a density operator $\rho$ representing the  collected electric field, the entries of the quantum FI matrix $\bm{\mathcal{K}}$ are given by \cite{helstrom1967minimum,helstrom1976quantum,braunstein1994statistical}
\begin{equation}
\label{eqn:QFI}
    \mathcal{K}_{ij}=\frac{1}{2}\Re{\Tr \rho\left(\mathcal{L}_{i}\mathcal{L}_{j}+\mathcal{L}_{j}\mathcal{L}_{i}\right)},
\end{equation}
where $\mathcal{L}_{i}$ is termed the symmetric logarithmic derivative (SLD) given implicitly by
\begin{equation}
\label{eqn:SLD}
    \pdv{\rho}{\mu_i}=\frac{1}{2}\left(\mathcal{L}_{i}\rho+\rho\mathcal{L}_{i}\right).
\end{equation}

Using a vectorial diffraction model \cite{bohmer2003orientation,lieb2004single,axelrod2012fluorescence,backer2014extending,backer2015determining,Chandler:19}, we express the wavefunctions of a photon emitted by a rotationally fixed molecule at the back focal plane (BFP) of the imaging system {[\Cref{fig:imagingSys_bfp}(a)]} as
\begin{linenomath} \begin{subequations}
\begin{align}
    \psi_x(u,v;\bm{\mu})&=[g_1(u,v),g_2(u,v),g_3(u,v)]{\cdot}\bm{\mu}\\
    \psi_y(u,v;\bm{\mu})&=[g_2(v,u),g_1(v,u),g_3(v,u)]{\cdot}\bm{\mu} \\
    \psi_z(u,v;\bm{\mu})&=0
\end{align}
\end{subequations} \end{linenomath}
where $(\psi_x,\psi_y,\psi_z)$ denote linearly polarized fields along $(x,y,z)$. The basis fields at the BFP of the imaging system $(g_1,g_2,g_3)$ may be interpreted as the classical electric field patterns produced by dipoles aligned with the $(x,y,z)$ Cartesian axes and projected by the microscope objective into the BFP [\Cref{appendix:1stOrder,eqn:basisFieldsBFP}].

To proceed in writing down the photon density operator $\rho$ collected by an objective lens, we define a scalar wavefunction
\begin{equation}
    \psi(u,v) = \psi_x(u,v)+\psi_y(u',v')
    \label{eqn:wavefxnDefn_psi}
\end{equation}
such that $x$- and $y$-polarized photons are detected separately and simultaneously, i.e., $[u',v']=[u-u_0,v-v_0]$ represents a translation $\sqrt{u_0^2+v_0^2}>r_0$ of $\psi_y$ (e.g., by a pair of mirrors) such that $\psi_x$ and $\psi_y$ are spatially resolvable. Here, the dimensionless scalar $r_0=\text{NA}/n$ represents the radius of the pupil of the imaging system (normalized by the focal length of the collection objective) as a function of the numerical aperture NA and the refractive index of the imaging medium $n$, which is assumed to be matched to that of the sample. Similarly, we define {[\Cref{fig:imagingSys_bfp}(b)]}
\begin{linenomath} \begin{subequations}
\label{eqn:defn_g}
\begin{align}
    g_x(u,v) &= g_1(u,v)+g_2(v',u') \label{eqn:defn_gx}\\
    g_y(u,v) &= g_2(u,v)+g_1(v',u') \label{eqn:defn_gy}\\
    g_z(u,v) & =g_3(u,v)+g_3(v',u') \label{eqn:defn_gz}
\end{align}
\end{subequations} \end{linenomath}
such that the wavefunction can be written as
\begin{equation}
    \psi(u,v) =g_x(u,v)\mu_x+g_y(u,v)\mu_y+g_z(u,v)\mu_z.
    \label{eqn:wavefxnDefn_g}
\end{equation}
Therefore, {if we neglect multiphoton events \cite{tsang2016quantum}, the zero- and} one-photon state can be represented by 
\begin{equation}
    \rho=(1-\epsilon_z)\op{\text{vac}}+\op{\psi},
\end{equation}
where $\ket{\text{vac}}$ denotes the vacuum state, where no photon is captured by the objective lens. Stemming from the finite NA of the imaging system, the probability of {detecting a photon emitted by the dipole is given by $\epsilon_z=\ip{\psi}=1-(1-c)\mu_z^2$,} where
\begin{equation}
    \ket{\psi}=\iint\psi(u,v)\ket{u,v}\,du\,dv
\end{equation}
and $\ket{u,v}$ denotes the position eigenket such that $\ip{u,v}{u',v'}=\delta(u-u')\delta(v-v')$. The scalar $c$ can be viewed as the probability of collecting a photon from a $z$-oriented molecule, normalized to that from an $x$- or $y$-oriented dipole, given by (\Cref{appendix:1stOrder})
\begin{equation}
\label{eqn:c}
    c=\ip{g_z}=\frac{-2r_0^4+6r_0^2-12\sqrt{1-r_0^2}+12}{r_0^4-9r_0^2+24} \in(0,1).
\end{equation}
Throughout this paper, we use $n=1.515$ and $\text{NA}=1.4$, i.e., $r_0=0.924$ and $c=0.65$, if not otherwise specified.
 
In \Cref{appendix:1stOrder}, we derive the QCRB for estimating the first-order orientational moments, yielding 
\begin{equation}
    \label{eqn:QFI_1st}
    \bm{\mathcal{K}}^{-1}=\frac{\mu_z^2}{F_p}\bm{\nu}_p\bm{\nu}_p^\dagger+\frac{1}{F_a}\bm{\nu}_a\bm{\nu}_a^\dagger,
\end{equation}
where eigenvectors $\bm{\nu}_p=[\mu_x,\mu_y]^\dagger/\sqrt{1-\mu_z^2}$ and $\bm{\nu}_a=[\mu_y,-\mu_x]^\dagger/\sqrt{1-\mu_z^2}$ represent orientational unit vectors along the polar and azimuthal directions, and
\begin{linenomath} \begin{subequations}
\begin{align}
    F_p &= 4(1+c-\epsilon_z) = 4c(1-\mu_z^2+\mu_z^2/c) \label{eqn:Fr} \\
    F_a& = 4 \label{eqn:Fa}
\end{align}
\end{subequations} \end{linenomath}
represent the QFI components along the polar and azimuthal directions, respectively. We may reparameterize this quantum limit in terms of the best-possible precision of measuring a dipole's orientation in polar coordinates $[\theta,\phi]$, given by
\begin{linenomath} \begin{subequations} \label{eqn:QCRB_1st} 
\begin{align}
    \sigma_{\theta,\text{QCRB}} &= \frac{1}{2\sqrt{c\,\sin^2\theta+\cos^2\theta}}\\
    \sigma_{\phi,\text{QCRB}} &= \frac{1}{2\sin\theta}.
\end{align}
\end{subequations} \end{linenomath}
Here, we use $\sigma_\text{QCRB}$ to denote the best-possible measurement standard deviation  for \emph{any} imaging system, as determined by the QFI, while we use $\sigma$ to denote the best-possible measurement standard deviation for a \emph{particular} imaging system, as determined by classical FI.

The QFI along the polar direction $F_p$ implicitly quantifies the change of the wavefunction $\psi$ with respect to the polar orientation of the source dipole and increases as $\mu_z$ increases (\Cref{eqn:Fr}). Given the toroidal emission pattern of a dipole, changes in polar orientation are easier to detect when sensing the null in the distribution (i.e., large $\mu_z$) in contrast to viewing the dipole from the side (i.e., large $\theta$). In the limiting case of $r_0=c=1$, the $4 \pi$ collection aperture captures the entire radiated field, and the limit of polar orientation precision $\sigma_{\theta,\text{QCRB}}$ becomes 0.5~rad for all possible orientations. 

Interestingly similar to estimating the 3D position of a dipole emitter \cite{backlund2018fundamental}, the QFI for measuring azimuthal orientation is uniform across all possible orientations [\Cref{eqn:Fa}], i.e., the best-possible uncertainty (as a longitudinal arc length on the orientation unit sphere) does not vary with NA or orientation $\bm{\mu}$. However, the circumference of the circles of latitude decrease with decreasing polar angle $\theta$, thereby causing the limit of azimuthal orientation precision $\sigma_{\phi,\text{QCRB}}$ to degrade as $1/(2 \sin\theta)$.

\begin{figure}[htbp]
    \centering
    \includegraphics{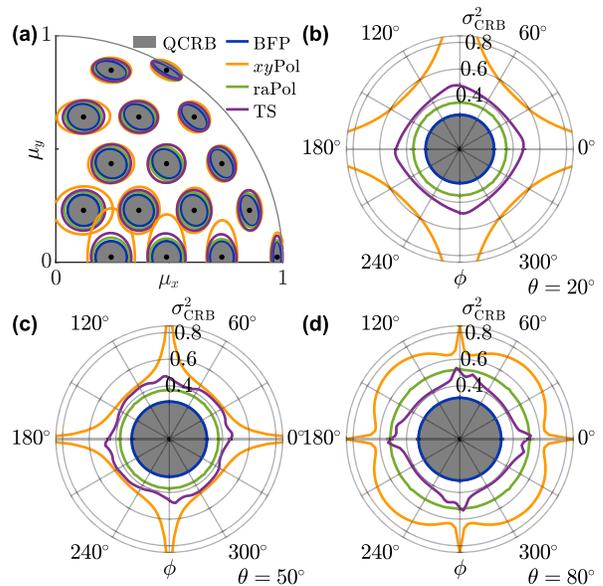}
    \caption{Classical CRB of several techniques (\Cref{appendix:classicalFI}) compared to the quantum CRB of estimating first-order orientational moments of fixed dipole emitters. (a) CRB covariance ellipses for measuring $[\mu_x,\mu_y]$ using 25 detected photons. {To compute the covariance for $N$ photons detected, scale the dimensions of the ellipses by $5/\sqrt{N}$.} (b\protect\nobreakdash-\hspace{0pt}d)~CRB standardized generalized variance (SGV in steradians) $\sigma_\text{CRB}^2=[\mu_z^2\,\det(\bm{\mathcal{J}})]^{-1/2}$ of estimating $[\mu_x,\mu_y]$, normalized to quantify orientation precision as a solid angle on the surface of a unit sphere. SGV is computed for a dipole with orientation (b)~$\theta=20^\circ$, (c)~$\theta=50^\circ$, and (d)~$\theta=80^\circ$ and 1 photon detected. For $N$ photons detected, SGV may be computed by scaling the radial axis by $1/N$. The gray regions are bounded from above by QCRB [\Cref{eqn:QFI_1st,eqn:QCRB_1st}]. Blue: direct BFP imaging (BFP), orange: standard PSF with $x$- and $y$-polarization separation ($xy$Pol), green: standard PSF with radial- and azimuthal-polarization separation (raPol), and purple: Tri-spot PSF with $x$- and $y$-polarized detection (TS).}
    \label{fig:QFI_FI_1st}
\end{figure}

We compare the classical CRB of multiple orientation measurement techniques to the quantum bound. Remarkably, direct BFP imaging (with $x$- and $y$-polarization separation) \cite{lieb2004single} has the best precision among the methods we compared, and since its variance ellipses overlap with the quantum bound, it achieves QCRB-limited measurement precision [\Cref{fig:QFI_FI_1st}(a)]. The widely used $x/y$-polarized standard PSF ($xy$Pol) \cite{mortensen2010optimized} has relatively poor precision compared to other techniques, as quantified by using standardized generalized variance $[\det(\bm{\mathcal{J}})]^{-1/2}$ (SGV), defined as the positive $p^\text{th}$ root of the determinant of a $p\times p$ covariance matrix \cite{sengupta1987tests}. SGV scales linearly with the area of the covariance ellipse for estimating $[\mu_x,\mu_y]$, and the SGV of the $xy$Pol technique is approximately three times larger on average than the quantum bound for out-of-plane molecules [\Cref{fig:QFI_FI_1st}(b)] and twice larger for in-plane molecules [\Cref{fig:QFI_FI_1st}(d)]. Its precision in measuring $x$- and $y$-oriented molecules is severely hampered due to its symmetry and resulting measurement degeneracy. The Tri-spot (TS) PSF, a PSF engineered specifically to measure molecular orientation \cite{zhang2018imaging}, has better overall precision compared to the $x/y$-polarized standard PSF, and its performance degrades only slightly for $x$- and $y$-oriented molecules. However, its precision does not reach the quantum limit. 

Note that both the $x/y$-polarized standard and TS PSFs break the azimuthal symmetry associated with conventional imaging systems, leading to $\phi$-dependent performance. Inspired to retain this symmetry, we also characterize the radially/azimuthally polarized version of the standard PSF (raPol) \cite{lew2014azimuthal}. This PSF is implemented by placing a vortex {(half)} wave plate (VWP), S-waveplate, or y-phi metasurface mask \cite{backlund2016removing} at the BFP. These elements convert radially  and azimuthally polarized light into linearly polarized light with orthogonal polarizations; these polarizations may be separated downstream by using a polarization beamsplitter (PBS). This technique has uniform precision for measuring molecular orientation across all azimuthal angles $\phi$ due to its symmetry. Its measurement precision is better than that of the TS PSF for most orientations [\Cref{fig:QFI_FI_1st}(b,c)] and only slightly worse for in-plane molecules [\Cref{fig:QFI_FI_1st}(d)].

\section{Reaching the Quantum Limit of Orientation Measurement Precision}
\label{sec:reachingQCRB}
Although direct BFP imaging achieves quantum-limited precision, it can only measure the orientation of one molecule at a time, thereby limiting its practical usage. In contrast, the aforementioned widefield imaging techniques can resolve the orientations of multiple molecules simultaneously, but their precisions do not reach the QCRB (\Cref{appendix:classicalFI}). Here, we analyze the classical FI of an imaging system [\Cref{eqn:classicalFI}] to deduce the conditions necessary for achieving the best-possible precision equal to the QCRB. 

The expected intensity distribution in the image plane is given by $I=\abs{U(\psi)}^2$, where $U$ is a unitary operator, i.e., {$\forall\,(\psi_1,\psi_2)$} $\ip{U(\psi_1)}{U(\psi_2)}=\ip{\psi_1}{\psi_2}$, that depends on the configuration of the imaging system. This linear operator $U$ typically involves a scaled Fourier transform ($xy$Pol), a Fourier transform after phase modulation (TS), or a Fourier transform after modulation by a polarization tensor (raPol). 
We consider an operator $U(\cdot)$ projecting the wavefunction $\psi(u,v)$ to the image plane such that the resulting field is either real or imaginary at any position $[u,v]$, i.e., the non-negative intensity is given by $I=[U(\psi)]^2$ or $I=-[U(\psi)]^2$. Therefore, \Cref{eqn:classicalFI} can be simplified to become
\begin{equation}
    \label{eqn:classicalFI_sim}
    \mathcal{J}_{ij}=\iint4 \left[\pdv{\mu_i}U(\psi)\right] \left[\pdv{\mu_j}U(\psi)\right]\,du\,dv.
\end{equation}
Further, since the basis fields remain mutually orthogonal after a unitary operation $U$, i.e., $\iint[U(g_i)][U(g_j)]\,du\,dv=0\ \forall\ i \neq j$, we find that the classical FI becomes equal to the QFI (\Cref{appendix:classicalFI}).

\begin{figure}[ht]
    \centering
    \includegraphics{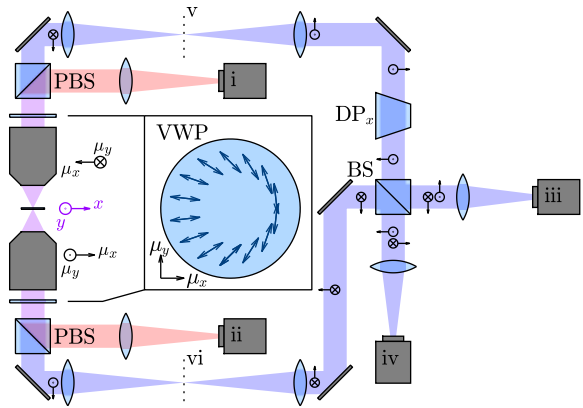}
    \caption{{Dual opposing-objective interferometric imaging (dualObj) for achieving QCRB-limited precision. Two vortex (half) waveplates (VWP) are placed at the BFPs to convert radially polarized to $x$-polarized light (blue) and azimuthally polarized to $y$-polarized light (red). Blue arrows depict the fast axis direction of the VWP. One of the radially polarized channels is flipped using an $x$-oriented Dove prism (DP) and then propagates to the beamsplitter (BS).}}
    \label{fig:dualObjSys}
\end{figure}

Therefore, an imaging system achieves the QFI limit for measuring dipole orientations if its images contain non-overlapping (i.e., non-interfering) real and imaginary fields. Further, in \Cref{appendix:classicalFI}, we find that the classical FI of a measurement saturates the quantum bound if and only if the phase of the detected electric field does not contain orientation information, i.e., $\abs{U(\psi)}\pdv*{\arg\{U(\psi)\}}{\mu_i}=0$. BFP imaging, where $U$ is the identity operator, satisfies this condition, and its precision reaches the quantum limit. In contrast, the field at the image plane is simply related to the field at the BFP by a Fourier transform; therefore, to satisfy the condition, a system may separate real and imaginary electric fields at the image plane, which is equivalent to separating even and odd field distributions at the BFP due to the parity of the Fourier transform. Alternatively, measuring the full complex field, i.e., both its amplitude and phase, could in principle reach the quantum limit of measurement precision.

\begin{figure}[htbp]
    \centering
    \includegraphics{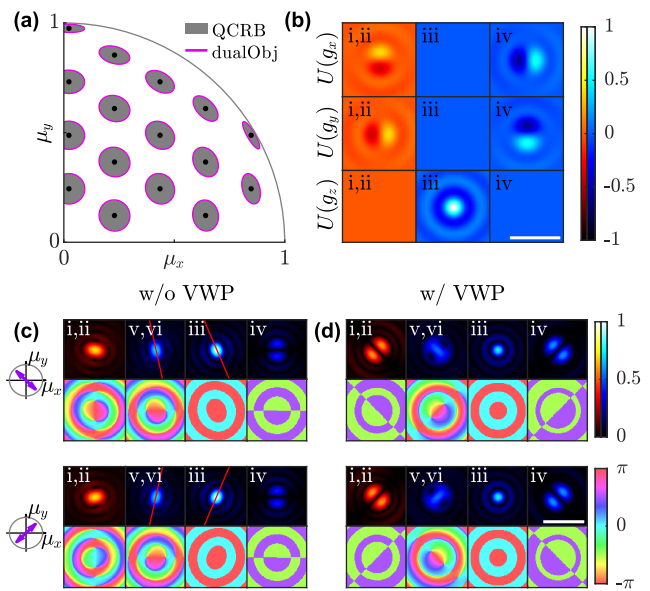}
    \caption{{Estimation precision and optical fields produced by the dual objective interferometric imaging system. (a) CRB covariance ellipses for measuring $[\mu_x,\mu_y]$ using 25 detected photons and interferometric detection (magenta) compared to the quantum bound. To compute the covariance for $N$ photons detected, scale the dimensions of the ellipses by $5/\sqrt{N}$. (b) Basis electric fields $U(g_x)$, $U(g_y)$ and $U(g_z)$ at detectors i-iv in \Cref{fig:dualObjSys}. (c-d) Normalized amplitude and phase of the optical fields of molecules with orientations $-\mu_x=\mu_y=\mu_z$ and $\mu_x=\mu_y=\mu_z$ captured at detectors i-iv and intermediate image planes v,vi in \Cref{fig:dualObjSys}(c) without and (d) with VWPs. Scale bar: 1 $\mathrm{\mu m}$. Colorbars: normalized amplitude and phase in rad.}}
    \label{fig:dualObj_FI_basis}
\end{figure}

Leveraging this insight, we propose an interferometric imaging system (dualObj, \Cref{fig:dualObjSys}) to measure the orientations of multiple molecules simultaneously with precision reaching the QCRB. This system uses two opposing objectives to collect the field emanated by a dipole, in a manner similar to 4Pi microscopy and iPALM {\cite{Hell:94,Shtengel3125,Aquino2011,Huang2016}}. To model the fields captured by each lens, we define orientation coordinates $(\mu_x,\mu_y)$ such that the two captured fields have identical amplitude distributions in the BFP, i.e., due to dipole symmetry, orientation coordinates $(\mu_x,\mu_y)$ are not the same as position coordinates $(x,y)$ as depicted in \Cref{fig:dualObjSys}. VWPs are placed at the BFPs to transform radially  and azimuthally polarized light into $x$- and $y$-polarized light, respectively.
Cameras (i) and (ii) detect identical images of the $y$ (azimuthally)-polarized fields. The $x$ (radially)-polarized fields, one of which is flipped by a dove prism (DP) (\Cref{fig:dualObjSys}), are guided to a beamsplitter (BS). The resulting interference pattern is captured by cameras (iii) and (iv). 

The precision of this interferometric imaging system saturates the QCRB [\Cref{fig:dualObj_FI_basis}(a)], since (1) the basis fields $U(g_x)$, $U(g_y)$, and $U(g_z)$ captured across cameras (i-iv) are mutually orthogonal [\Cref{fig:dualObj_FI_basis}(b)], and (2) the real [\Cref{fig:dualObj_FI_basis}(b)(i,ii,iv)] and imaginary [\Cref{fig:dualObj_FI_basis}(b)(iii)] components of the field are spatially separated. QCRB-limited precision can also be achieved by using a single objective and a 50/50 beamsplitter, as shown in \Cref{fig:singleObj_FI_basis}, but this system cannot measure positions and orientations of molecules simultaneously (\Cref{appendix:oneObj}). Note that although the photon detection rate is doubled in experiments using dual-objective detection, the two schemes exhibit identical orientational precision per photon detected.

To demonstrate the features of this optical design, we consider the optical fields of molecules with orientations $\bm{\mu}=[-1,1,1]/\sqrt{3}$ and $\bm{\mu}=[1,1,1]/\sqrt{3}$, propagated by the proposed imaging system to the various image planes [\Cref{fig:dualObj_FI_basis}(c,d)]. Corresponding images with Poisson shot noise are shown in \Cref{fig:singleObj_FI_basis}(b,c). Without including the VWP, the fields at cameras (i,ii) and intermediate image planes (IIPs, v,vi) represent the response of an $x/y$-polarized imaging system [\Cref{fig:dualObj_FI_basis}(c)(i,ii,v,vi)]. Both the amplitudes and phases of the fields contain orientation information, but the phase patterns are lost when using photon-counting cameras. Therefore, the performance of the $xy$Pol imaging system is worse than the quantum bound. 
After guiding the $y$-polarized fields to the interferometric detection path, the phase shift induced by the BS separates the real and imaginary fields, i.e., the phase patterns of the fields detected are binary [\Cref{fig:dualObj_FI_basis}(c)(iii,iv)] and do not contain orientation information. Images of these two dipoles are now easier to distinguish from one another, as exemplified by rotation in the elongated PSFs [red lines in \Cref{fig:dualObj_FI_basis}(c)(iii) vs. \Cref{fig:dualObj_FI_basis}(c)(v,vi)].

While interferometric detection can also be implemented in the $x$-polarized channel [\Cref{fig:dualObj_FI_basis}(c)(i,ii)] to boost precision, we notice that a VWP combined with a PBS separates radially  and azimuthally polarized light, and all basis electric fields in the azimuthal channel are odd at the BFP, i.e., the basis fields are completely imaginary in the image plane [\Cref{fig:dualObj_FI_basis}(d)(i,ii)]. Therefore, using a VWP eliminates the need for interferometric detection in the azimuthal channel, yielding a simpler imaging system. In the radially polarized channel [\Cref{fig:dualObj_FI_basis}(d)(v,vi)], we implement interferometric detection to improve image contrast [\Cref{fig:dualObj_FI_basis}(d)(iii,iv)], thereby enabling QCRB-limited orientation measurement precision [\Cref{fig:dualObj_FI_basis}(a)]. {Further, this imaging system also saturates the QCRB for measuring the 3D position of SMs \cite{backlund2018fundamental}, making it optimal for both 3D orientation and 3D position measurements. While complex to implement and align, the required polarization elements can be added directly to existing dual-objective imaging systems \cite{Aquino2011,Huang2016}.}

\section{Fundamental Limits of Measuring Orientation and Wobble Simultaneously}
\begin{figure}[htbp]
    \centering
    \includegraphics{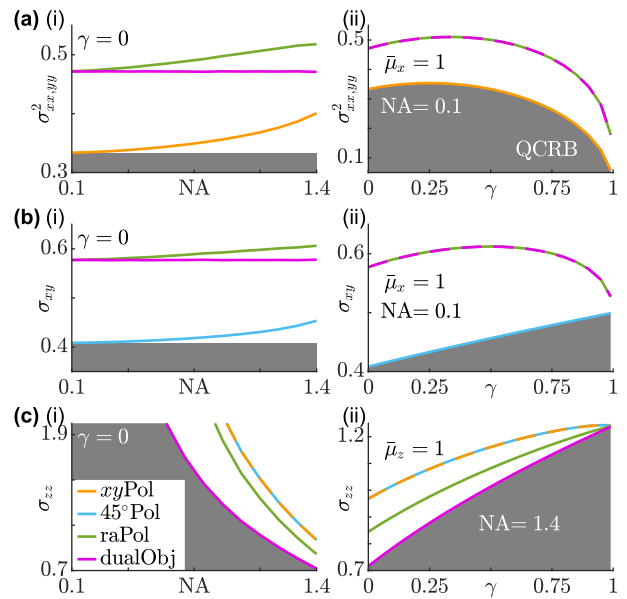}
    \caption{Classical CRB of several techniques (\Cref{appendix:2ndOrderClassical}) compared to the quantum CRB of estimating second-order orientational moments of dipole emitters. (a) CRB SGV of estimating $M_{xx}$ and $M_{yy}$ for molecules wobbling around the $\mu_x$ axis, (b-c) best-possible precision $\sqrt{\text{CRB}}$ of estimating (b)~$M_{xy}$ for molecules wobbling around the $\mu_x$ axis and (c) $M_{zz}$ for molecules wobbling around the $\mu_z$ axis as functions of (i) numerical aperture NA (for $\gamma=0$) and (ii) rotational constraint $\gamma$ [$\text{NA}=0.1$ in (a),(b) and $\text{NA}=1.4$ in (c)]. The gray regions are bounded from above by (a)~QCRB or (b,c)~$\sqrt{\text{QCRB}}$ [\Cref{eqn:QFI2nd}]. Orange: standard PSF with $x$- and $y$-polarized detection ($xy$Pol), cyan: standard PSF with linearly polarized detection at $\pm 45^\circ$ in the $xy$-plane ($45^\circ$Pol), green: standard PSF with radially  and azimuthally polarized detection (raPol), and magenta: dual-objective interferometric detection with VWPs (dualObj). All curves assume 1 photon is detected from the dipole emitter. {For $N$ photons detected, scale the vertical axis values by (a) $1/N$ and (b,c)~$1/\sqrt{N}$.} The estimation precision of $45^\circ$Pol in (a) and $xy$Pol in (b) are orders of magnitude larger than those of the other techniques and are not shown.}
    \label{fig:QCRB_CRB_2nd_NA_gamma}
\end{figure}
While a single photon emitted by a dipole has a wavefunction $\psi$ that is consistent with a single orientation $\bm{\mu}$, camera images usually contain multiple photons, thereby inherently enabling measurements of rotational dynamics during a camera's integration time \cite{backer2016enhanced,zhang2018imaging,Mazidi2019,curcio2019birefringent}. Note that a collection of photons emitted by a partially fixed or freely rotating molecule is equivalent to that emitted by some collection of fixed dipoles with a corresponding orientation distribution. Therefore, the photon state for a wobbling molecule may be expressed as a mixed state density matrix
\begin{linenomath} \begin{align}
    \rho&=\frac{1}{T}\int_0^T\rho[\bm{\mu}(t)]\,dt\nonumber\\
    &=(1-c)M_{zz}\op{\text{vac}}+\sum_{i,j\in \{x,y,z\} }\op{g_i}{g_j}M_{ij},
\end{align} \end{linenomath} 
where $M_{ij}=(1/T)\int_0^T\mu_i\mu_j\,dt$ is the temporal average of the second moments of molecular orientation over acquisition time $T$. The corresponding classical image formation model is given by \Cref{eqn:classical_image_formation_2nd,eqn:basis_images_classical}. The QFI may be expressed as a function of the orientational second moments and can be computed numerically as shown in \Cref{appendix:secondOrder}. 

For simplicity, we parameterize a dipole's rotational motions by using an average orientation $[\bar{\mu}_x,\bar{\mu}_y,\bar{\mu}_z]$ with rotational constraint $\gamma$ \cite{zhang2018imaging,curcio2019birefringent,zhang2019fundamental}, i.e., 
\begin{linenomath} \begin{subequations}
    \label{eqn:M_vs_muGamma} 
    \begin{align}
        M_{ii} &= \gamma\bar{\mu}_i^2+\frac{1-\gamma}{3} & i\in\{x,y,z\}\\
        M_{ij} &= \gamma\bar{\mu}_i\bar{\mu}_j &i,j\in\{x,y,z\},\,i\neq j,
    \end{align}
\end{subequations} \end{linenomath}
where $\gamma=0$ represents a freely rotating molecule and $\gamma=1$ indicates a rotationally fixed molecule. We may derive an analytical expression of QFI for estimating a subset of the second moments $[M_{xx},M_{yy},M_{zz},M_{xy}]$ (\Cref{appendix:secondOrder}) by examining special cases where the dipole's average orientation is parallel to the Cartesian axes. The QFI matrices for a dipole with an average orientation along the $x$ axis [$\bm{\mathcal{K}}^x$, i.e., $\bar{\mu}_x=1,M_{xy}=M_{xz}=M_{yz}=0$, \Cref{fig:QCRB_CRB_2nd_NA_gamma}(a,b)] and that for a dipole with an average orientation parallel to the optical axis $\mu_z$ [$\bm{\mathcal{K}}^z$, i.e., $\bar{\mu}_z=1,M_{xy}=M_{xz}=M_{yz}=0$, \Cref{fig:QCRB_CRB_2nd_NA_gamma}(c)] are given by
\begin{linenomath} \begin{subequations}
    \label{eqn:QFI2nd}
    \begin{align}
    \bm{\mathcal{K}}^x&=\,\text{diag}\left(\mathcal{K}_{xx}^x,\mathcal{K}_{yy}^x,\mathcal{K}_{zz}^x,\mathcal{K}_{xy}^x\right)\nonumber \\
    &=\,\text{diag}\left(\frac{3}{1+2\gamma},\frac{3}{1-\gamma},\frac{3c}{1-\gamma},\frac{12}{2+\gamma}\right) \label{eqn:QFI2nd_x}\\
    \bm{\mathcal{K}}^z&=\,\text{diag}\left(\frac{3}{1-\gamma},\frac{3}{1-\gamma},\frac{3c}{1+2\gamma},\frac{6}{1-\gamma}\right). \label{eqn:QFI2nd_z}
    \end{align}
\end{subequations} \end{linenomath}

One sufficient condition to saturate the QFI for estimating a subset of parameters is for the measurement to project onto the eigenstates of the corresponding SLDs \cite{braunstein1994statistical}. For example, when a low NA objective lens is used, the $x/y$-polarized standard PSF separates nearly perfectly the basis images corresponding to $M_{xx}$ and $M_{yy}$ and has no sensitivity to $M_{zz}$. Therefore, the $x/y$-polarized standard PSF projects onto the eigenstates of $\mathcal{L}_{xx}$ and $\mathcal{L}_{yy}$ and its precision approaches the QCRB limit for measuring $M_{xx}$ and $M_{yy}$ for small NA [\Cref{fig:QCRB_CRB_2nd_NA_gamma}(a)]. However, this technique lacks sensitivity for measuring the cross moment $M_{xy}$ [\Cref{fig:QCRB_CRB_2nd_NA_gamma}(b)] since the corresponding FI entry is close to zero [\Cref{fig:QFI_FI_2nd_NA_gamma}(b)]. Intuitively, $M_{xy}$ may be measured simply by rotating the polarizing beamsplitter by $45^\circ$ around the optical axis to capture linearly polarized light along $\pm 45^\circ$. This approach achieves the QFI limit for measuring $M_{xy}$, but consequently contains no information regarding the squared moments $M_{xx}$ and $M_{yy}$ [\Cref{fig:QCRB_CRB_2nd_NA_gamma}(a,b), \Cref{fig:QFI_FI_2nd_NA_gamma}(a,b)].

To quantify measurement performance corresponding to out-of-plane second moments, we focus on the CRB $\sigma_{zz}$, since all polarized versions of the standard PSF have poor sensitivity for measuring cross moments $M_{xz}$ and $M_{yz}$. Not surprisingly, the precision of measuring $M_{zz}$ dramatically improves when using an objective lens of NA greater than 1 [\Cref{fig:QCRB_CRB_2nd_NA_gamma}(c)(i)]. Here, we notice the usefulness of dual-objective interferometric detection (dualObj); since the photons corresponding to $M_{zz}$ are separated from other second moments [\Cref{fig:dualObj_FI_basis}(b)], i.e., the system projects onto the eigenstate of $\mathcal{L}_{M_{zz}}$, dualObj achieves QCRB-limited precision [\Cref{fig:QCRB_CRB_2nd_NA_gamma}(c)]. Without interferometric detection (raPol), radially/azimuthally polarized detection achieves worse precision than dualObj but improves upon basic linear polarization separation ($xy$Pol or $45^\circ$Pol). 

Close examination of \Cref{fig:QCRB_CRB_2nd_NA_gamma}(a,b) shows that no existing orientation imaging methods, even those that achieve QCRB-limited precision for estimating first moments, can achieve QFI-limited precision for measuring all orientational second moments simultaneously. To gain insight into this phenomenon, we use classical FI to analyze the SGV $\left(\sigma^{i}_{xx,yy,xy}\right)^2$  of measuring all in-plane moments simultaneously (\Cref{appendix:2ndOrderClassical}), yielding
\begin{linenomath} \begin{multline}
\label{eqn:boundTradeOff}
    \left(\sigma^{i}_{xx,yy,xy}\right)^2 = \left[\text{det}\left(\bm{\mathcal{J}}_{xx,yy,xy}^i\right)\right]^{-1/3} \\
    \geq\,\left(\mathcal{K}_{xx}^i\mathcal{K}_{yy}^i\mathcal{K}_{xy}^i/4\right)^{-1/3}=\left[\text{det}\left(\bm{\mathcal{K}}_{xx,yy,xy}^i\right)/4\right]^{-1/3},
\end{multline} \end{linenomath}
where the superscript $i \in \{x,y,z\}$ denotes the SGV $\sigma^2$, FI $\bm{\mathcal{J}}$, or QFI $\bm{\mathcal{K}}$ of a dipole with an average orientation along one of the Cartesian axes. 
\begin{figure}[htbp]
    \centering
    \includegraphics{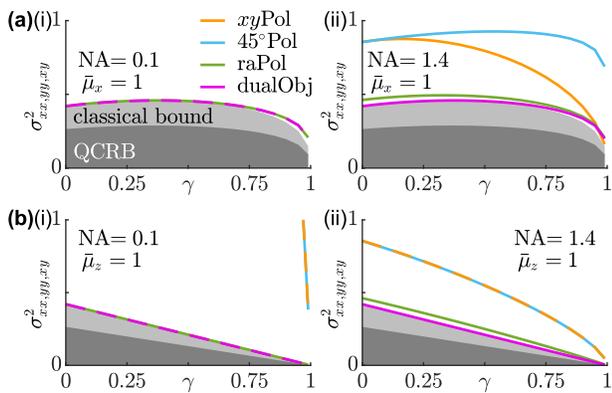}
    \caption{CRB standardized generalized variance (SGV in steradians) of estimating in-plane moments $M_{xx}$, $M_{yy}$, and $M_{xy}$ simultaneously for molecules wobbling around the (a)~$\mu_x$ axis and (b) $\mu_z$ axis using (i) $\text{NA}=0.1$ and (ii) $\text{NA}=1.4$ objective lenses. The dark gray regions are bounded from above by the quantum bound [\Cref{eqn:QFI2nd}]; light gray regions are bounded from above by the classical bound [\Cref{eqn:boundTradeOff}]. Orange: standard PSF with $x$- and $y$-polarized detection ($xy$Pol), cyan: standard PSF with linearly polarized detection at $\pm 45^\circ$ in the $xy$-plane ($45^\circ$Pol), green: standard PSF with radially  and azimuthally polarized detection (raPol), and magenta: dual-objective interferometric detection with VWPs (dualObj). All curves assume 1 photon is detected from the dipole emitter. {For $N$ photons detected, scale the vertical axis values by $1/N$.} The estimation precision of $45^\circ$Pol and $xy$Pol in (i) are orders of magnitude larger than those of the other techniques and are not shown.}
    \label{fig:classicalCRB_tradeoff}
\end{figure}

\Cref{eqn:boundTradeOff} reveals that there exists a trade-off between sensitivity for measuring squared moments, which mainly indicate the average orientation of a molecule, versus cross moments, which correspond to wobble [\Cref{eqn:M_vs_muGamma}], for all imaging systems. Radially/azimuthally polarized standard PSFs, both with (dualObj) and without (raPol) interferometric detection, exhibit nearly identical precision for measuring squared moments versys cross moments [\Cref{fig:QFI_FI_2nd_NA_gamma}(a,b)] and perform closely to the bound given by \Cref{eqn:boundTradeOff} for both low [\Cref{fig:classicalCRB_tradeoff}(i)] and high NA [\Cref{fig:classicalCRB_tradeoff}(ii)]. In contrast, the linearly polarized standard PSFs, $xy$Pol and raPol, exhibit suboptimal SGVs $\sigma_{xx,yy,xy}^2$ for measuring all in-plane second moments simultaneously for low NA as expected [\Cref{fig:classicalCRB_tradeoff}(i)], and these SGVs improve as NA increases [\Cref{fig:classicalCRB_tradeoff}(ii)]. This improvement comes at the cost of worsening measurement precision for specific moments [$(M_{xx},M_{yy})$ for $xy$Pol and $M_{xy}$ for $45^\circ$Pol, \Cref{fig:QCRB_CRB_2nd_NA_gamma}(a,b)(i)]. Interestingly, no method can achieve QCRB-limited measurement precision for all second-order orientational moments simultaneously since the bound given by \Cref{eqn:boundTradeOff} is greater than the quantum bound [\Cref{eqn:QFI2nd}]. This trade-off also occurs for molecules wobbling around other average orientations (\Cref{fig:QCRB_CRB_2nd_fixed}).

\section{Discussion and Conclusion}
Using quantum estimation theory, we derive a fundamental bound for estimating the orientation of rotationally fixed molecules that applies to all measurement techniques. The key result is that the bound is radially symmetric; the precision along the polar direction depends on the numerical aperture of the imaging system and the polar orientation $\mu_z$ of the molecule, while the precision along the azimuthal direction is bounded by a constant 0.5 rad. Our approach can be extended to include appropriately modeled background photons (\Cref{appendix:background}). {Estimation performance can vary dramatically depending on how the background photons interact with the signal photons and the parameters to be estimated, and exploring these effects for typical single-molecule imaging conditions remains the object of future study.} By comparing the precision of existing methods to the bound, we show that direct imaging of the BFP saturates the quantum bound, while all existing image-plane based techniques have worse precision. Upon further investigation of the classical FI, we show that a method can saturate the quantum bound if and only if the field in the image plane contains only trivial phase information. Inspired by this necessary and sufficient condition, we propose an imaging system with interferometric detection at the image plane that saturates the quantum bound.

We further examined the quantum bound for estimating the {average} orientation and wobble of a non-fixed molecule. {Since the orientation and wobble measurement is composed of a number
of individual molecular orientations mixed together,} our analysis shows that the optimality of a measurement depends on the specific molecular orientation {trajectory} to be observed. Although no measurement is physically realizable that achieves QCRB-limited precision for all second moments and all possible molecular orientations simultaneously, we show several methods that achieve quantum-limited precision for certain subsets of second moments. Generally speaking, spatially separating basis fields improves the precision of measuring the average orientation of an SM, while mixing (i.e., increasing the spatial overlap of the) basis fields improves the precision of measuring their wobble. The trade-off is demonstrated using classical FI (\Cref{appendix:2ndOrderClassical}). An imaging system that separates radially  and azimuthally polarized light using a VWP and a PBS is capable of distributing information evenly between measuring the average orientation and wobble (raPol and dualObj in \Cref{fig:QCRB_CRB_2nd_NA_gamma}), and these methods achieve optimal measurement precision for in-plane moments in terms of CRB SGV [\Cref{fig:classicalCRB_tradeoff}]. Although we model the orientation of SMs using orientational second-order moments, similar results can also be derived for other orientation parameterizations such as generalized Stokes vectors and spherical harmonics (Appendix \ref{appendix:secondOrder}). 

Interestingly, we note that certain entries of the QFI matrix may be infinite, e.g., $\mathcal{K}^x_{yy,yy}=\mathcal{K}^x_{zz,zz}=\infty$ for fixed molecules oriented along the $x$ axis [\Cref{eqn:QFI2nd_x}] and $\mathcal{K}^z_{xx,xx}=\mathcal{K}^z_{yy,yy}=\mathcal{K}^z_{xy,xy}=\infty$ for fixed molecules oriented along the $z$ axis [\Cref{eqn:QFI2nd_z}]. Such cases arise when $\rho\pdv*{\rho}{M_{ij}}$ vanishes as a molecule becomes more fixed ($\gamma \rightarrow 1$). One such example is using the $x$/$y$-polarized standard PSF to estimate $M_{yy}$ for an $x$-oriented fixed molecule; the classical FI $\mathcal{J}^x_{yy,yy}$ is also infinite in this case. That is, there exists some position(s) $(u,v)$ in image space such that $I(u,v;\bm{\mu}=[1,0,0]^\dagger)=0$ and $\pdv*{I(u,v)}{M_{yy}}>0$, i.e., we expect certain region(s) of the image to be dark for $x$-oriented dipoles but bright for $y$-oriented dipoles. Therefore,
\begin{equation}
    \mathcal{J}^x_{yy,yy} =\iint \frac{\left[\pdv*{I(u,v)}{M_{yy}}\right]^2}{I(u,v)}\,du\,dv=\infty
\end{equation}
This situation is the orientation analog of MINFLUX nanoscopy \cite{Balzarotti606}, where infinitely good orientation measurement precision per photon may be obtained by receiving zero signal \cite{Paur:19}; in this case, zero photons detected in the $y$-polarized channel implies $M_{yy}=0$. 

It is remarkable that quantum estimation theory provides fundamental bounds on measurement performance that are both instrument-independent and achievable by readily built imaging systems, such as the dual-objective system with vortex waveplates and interferometric detection proposed here. Further, these bounds give tremendous insight to microscopists, who can now compare existing methods for measuring dipole orientation to the bound and design new microscopes that optimally utilize each detected photon for maximum measurement precision. In particular, our analysis reveals that no single instrument can achieve the best-possible QCRB limit for measuring all orientational second moments simultaneously due to the trade-off between measuring mean orientation versus molecular wobble [\Cref{eqn:boundTradeOff}]. Therefore, the notion of designing a single, fixed instrument that performs optimally may simply be intractable, and instead, scientists and engineers should focus on designing ``smart'' imaging systems that adapt to the specific dipole orientations within the sample and orientational second moments of interest, thus achieving optimal, QFI-limited measurement precision. Such designs remain the object of future studies.

\begin{acknowledgments}
We acknowledge the helpful discussions with Tianben Ding, Tingting Wu, Hesam Mazidi, and Dr. Jin Lu. This work was supported by the National Science Foundation under grant number ECCS-1653777 and by the National Institute of General Medical Sciences of the National Institutes of Health under grant number R35GM124858.
\end{acknowledgments}

\appendix

\section{Quantum Fisher information of estimating first-order orientational dipole moments}
\label{appendix:1stOrder}
\begin{figure}[h]
    \centering
    \includegraphics{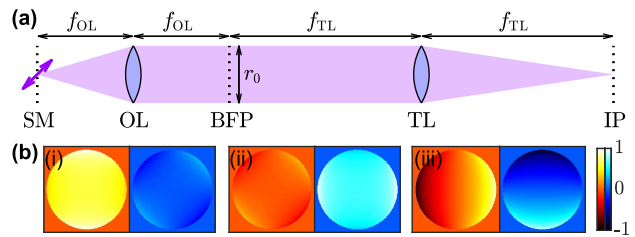}
    \caption{{Overview of a simplified two-lens imaging system. (a) An objective lens (OL) captures light emitted by a single-molecule (SM) emitter positioned at its focal plane. A tube lens (TL) is used to focus light to the image plane (IP). (b)~Basis electric fields (i) $g_x$, (ii) $g_y$, and (iii) $g_z$ at the back focal plane (BFP) of the imaging system normalized to the maximum amplitude of $g_x$. Red: $x$-polarized field; blue: $y$-polarized field.}}
    \label{fig:imagingSys_bfp}
\end{figure}

Here, we derive the Quantum Fisher information (QFI) of estimating the first-order orientational moments of a fixed dipole emitter. We consider the classical wavefunction given in \Cref{eqn:wavefxnDefn_g,eqn:defn_g}. The basis fields, as measured in the BFP, of each dipole moment are given by \cite{backer2014extending,Chandler:19}
\begin{linenomath} \begin{subequations}
\label{eqn:basisFieldsBFP}
\begin{align}
    g_1(u,v) &= A \Circ\left(\frac{r}{r_0}\right)\frac{u^2\sqrt{1-r^2}+v^2}{r^2(1-r^2)^{1/4}}
    \\
    g_2(u,v) &= A \Circ\left(\frac{r}{r_0}\right)\frac{uv\left(\sqrt{1-r^2}-1\right)}{r^2(1-r^2)^{1/4}}\\
    g_3(u,v) &= -A \Circ\left(\frac{r}{r_0}\right)\frac{u}{(1-r^2)^{1/4}},
\end{align}
\end{subequations} \end{linenomath}
where $r=\sqrt{u^2+v^2}${, and $\Circ(r)$ is an indicator function representing a circular aperture that equals 1 for $r \leq 1$ and 0 otherwise}. The scalar $A$ represents a normalization factor such that $\iint(g_1^2+g_2^2)\,du\,dv=1$ and $\iint g_3^2\,du\,dv=c/2$ [\Cref{eqn:c}], given by
\begin{equation}
    A^{-2}=\frac{\pi}{3}\left[4+\left(r_0^2-4\right)\sqrt{1-r_0^2}\right].
\end{equation}
Therefore, the photon state corresponding to wavefunction $\psi$ is given by $\rho=(1-\epsilon_z)\op{\text{vac}}+\op{\psi}$, where
\begin{equation}
    \ket{\psi}=\ket{g_x}\mu_x+\ket{g_y}\mu_y+\ket{g_z}\mu_z
\end{equation}
and $(g_x, g_y, g_z)$ defined in \Cref{eqn:defn_g} {(\Cref{fig:imagingSys_bfp})}.

We first derive the symmetric logarithmic derivatives (SLDs) for measuring these parameters. The SLDs are given implicitly by \Cref{eqn:SLD}, where $\pdv*{\rho}{\mu_k}=\op{\psi}{\pdv*{\psi}{\mu_k}}+\op{\pdv*{\psi}{\mu_k}}{\psi}$. The partial derivatives of state vector $\ket{\psi}$ are given by
\begin{linenomath} \begin{subequations}
\begin{align}
    \ket{\pdv{\mu_x}\psi} &= \ket{g_x}-\ket{g_z}\frac{\mu_x}{\mu_z}\\
    \ket{\pdv{\mu_y}\psi} &= \ket{g_y}-\ket{g_z}\frac{\mu_y}{\mu_z},
\end{align}
\end{subequations} \end{linenomath}
where we have applied the constraint $\mu_x^2+\mu_y^2+\mu_z^2=1$. Further, we may perform eigendecomposition on the density matrix $\rho$ such that
\begin{equation}
    \rho=\sum_iD_i\op{e_i},
\end{equation}
with $\{D_i\}$ and $\{\ket{e_i}\}$ being its eigenvalues and eigenstates, respectively. The SLDs are therefore given explicitly by \cite{tsang2016quantum}
\begin{equation}
\label{eqn:sldexplicit}
    \mathcal{L}_{k}=\sum_{D_i+D_j\neq0}\frac{2}{D_i+D_j}\matrixel{e_i}{\pdv*{\rho}{\mu_k}}{e_j}\op{e_i}{e_j}.
\end{equation}
Eigenstates $\ket{e_i}$ contribute to the sum only if $\pdv*{\rho}{\mu_k}\ket{e_i}\neq0$. We find three eigenstates that contribute to the sum such that $\{\ket{e_1},\ket{e_2},\ket{e_3}\}$ span $\left\{\ket{\psi},\ket{\pdv*{\psi}{\mu_x}},\ket{\pdv*{\psi}{\mu_y}}\right\}$:
\begin{linenomath} \begin{subequations}
\label{eqn:eigenstates}
\begin{align}
    \ket{e_1} &=\frac{1}{\sqrt{\epsilon_z}} \ket{\psi}\\
    \ket{e_2} &= \frac{1}{\sqrt{1-\mu_z^2}}\left(\mu_y\ket{g_x}-\mu_x\ket{g_y}\right)\\
    \ket{e_3} &=\sqrt{ \frac{c}{\epsilon_z(1-\mu_z^2)}} \Big[ \mu_x\mu_z\ket{g_x}+\mu_y\mu_z\ket{g_y}\nonumber\\
    &\qquad\qquad\qquad\qquad+(\mu_z^2-1)c^{-1}\ket{g_z} \Big]
\end{align}
\end{subequations} \end{linenomath}
with corresponding eigenvalues $D_1=\epsilon_z$ and $D_2=D_3=0$. The SLDs are computed by substituting these eigenstates [\Cref{eqn:eigenstates}] and eigenvalues into \Cref{eqn:sldexplicit}. The elements of the QFI can be computed according to \Cref{eqn:QFI}, yielding the QFI matrix
\begin{equation}
\label{eqn:QFIappendix}
    \bm{\mathcal{K}}=\frac{4}{\mu_z^2}
    \begin{bmatrix}c\mu_x^2+\mu_z^2 & c\mu_x\mu_y \\ 
    c\mu_x\mu_y & c\mu_y^2+\mu_z^2\end{bmatrix}.
\end{equation}

{Note that this derivation depends solely on the orthogonality between the basis fields $\ket{g_x}$, $\ket{g_y},$ and $\ket{g_z}$. Therefore, \Cref{eqn:QFIappendix} may also be used if the sample's refractive index differs from that of the imaging medium; in this case, the constant $c$ is no longer given by \Cref{eqn:c} and would need to be adjusted accordingly.}

\section{Classical Fisher information of estimating first-order orientational dipole moments}
\label{appendix:classicalFI}
We evaluate the classical FI given by \Cref{eqn:classicalFI}. For simplicity, we write the field at the camera plane as $\Psi(u,v)=U(\psi)=A_\Psi(u,v)\exp[j\alpha_\Psi(u,v)]$ where $\left\{A_\Psi,\alpha_\Psi\right\} \in \mathbb{R}^2$ and $U(\cdot)$ is a unitary operator, such as a Fourier transform. We consider the diagonal entries of the classical FI given by [\Cref{eqn:classicalFI}]
\begin{linenomath} \begin{align}
    \mathcal{J}_{ii}&=\iint\left[\left(\frac{\Psi}{\Psi^*}\right)^{\!1/2}\pdv{\Psi^*}{\mu_i}+\left(\frac{\Psi^*}{\Psi}\right)^{\!1/2}\pdv{\Psi}{\mu_i}\right]^2 du\,dv \nonumber \\ 
    &= 4\iint \Re{\left(\frac{\Psi}{\Psi^*}\right)^{\!1/2}\pdv{\Psi^*}{\mu_i}}^2 du\,dv \nonumber \\
    &= 4\iint \left[\pdv{\mu_i}A_\Psi(u,v)\right]^2 du\,dv.
\end{align} \end{linenomath} 
Thus, because the intensity $\abs{\Psi(u,v)}^2$ is detected by a camera, orientation information is only useful if it is encoded within the field amplitude $A_\Psi(u,v)$, i.e., FI increases as $\pdv*{A_\Psi(u,v)}{\mu_i}$ increases. Any information that may be present within phase variations that arise from changes in orientation, given by $\pdv*{\alpha_\Psi(u,v)}{\mu_i}$, are simply lost and do not improve Fisher information. 

Both the field and its partial derivatives can be viewed as superpositions of image-plane basis fields
\begin{equation}
    G_i=U(g_i), \label{eqn:basisFieldsImg}
\end{equation}
analogous to the fields at the BFP [\Cref{eqn:basisFieldsBFP}], given by
\begin{linenomath} \begin{subequations}
\label{eqn:classicalFieldDer}
\begin{align}
    \Psi&=G_x\mu_x+G_y\mu_y+G_z\mu_z\\
    \pdv{\Psi}{\mu_x}&=G_x-G_z\frac{\mu_x}{\mu_z}\\
    \pdv{\Psi}{\mu_y}&=G_y-G_z\frac{\mu_y}{\mu_z}.
\end{align}
\end{subequations} \end{linenomath}
Interestingly, we find that
\begin{linenomath} \begin{align}
    \mathcal{J}_{ii} &\leq 4\iint \abs{\pdv{\mu_i}\Psi(u,v)}^2\,du\,dv \nonumber \\
    &=4\iint\abs{G_i-G_z\frac{\mu_i}{\mu_z}}^2 du\,dv \nonumber \\
    &=4\iint\left(\abs{G_i}^2+\frac{\mu_i^2}{\mu_z^2}\abs{G_z}^2\right)du\,dv \nonumber \\
    &= \bm{\mathcal{K}}_{ii}.
\end{align} \end{linenomath} 
with equality if and only if
\begin{linenomath} \begin{align}
    &\Im{\left[\frac{\Psi(u,v)}{\Psi^*(u,v)}\right]^{1/2}\pdv{\mu_i}\Psi^*(u,v)}\nonumber\\
    =\,&\Im{\pdv{\mu_i}A_\Psi(u,v)-jA_\Psi(u,v)\pdv{\mu_i}\alpha_\Psi(u,v)}\nonumber\\
    =\,&-A_\Psi(u,v)\pdv{\mu_i}\alpha_\Psi(u,v) = 0. \label{eqn:imgPlanePhaseChange}
\end{align} \end{linenomath} 
Thus, the classical FI may equal the quantum bound if and only if the phase of the image-plane field, $\alpha_\Psi(u,v)$, is constant as the dipole changes orientation. That is, if one can design an imaging system such that all changes in orientation correspond solely to changes in the image-plane field amplitude $A_\Psi(u,v)$, such an imaging system may achieve quantum-limited orientation measurement precision.

\section{Single-objective interferometric imaging system that reaches the quantum limit of measurement precision}
\label{appendix:oneObj}
\begin{figure}[htbp]
    \centering
    \includegraphics{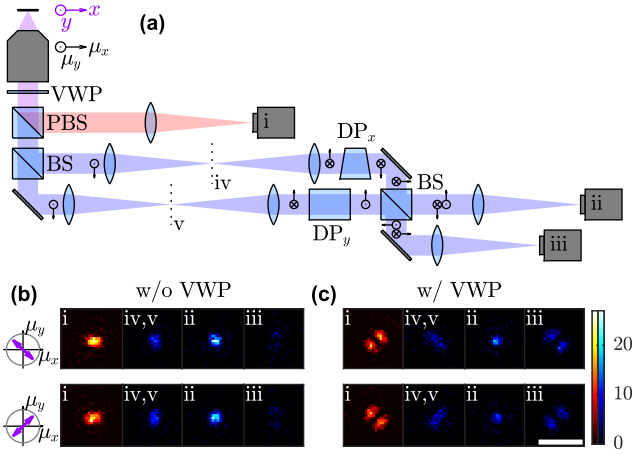}
    \caption{(a) A single-objective interferometric imaging system that reaches the quantum limit of measurement precision. A vortex waveplate (VWP) is placed at the BFP to convert radially  and azimuthally polarized light to $x$- (blue) and $y$-polarized (red) light, respectively, which is then separated by a polarizing beamsplitter (PBS). Camera (i) detects an azimuthally polarized image identical to those captured by cameras (i,ii) in \Cref{fig:dualObjSys}. The radial channel is split and recombined by a pair of 50/50 beamsplitters (BS) in a Mach-Zehnder configuration; light in each arm is flipped using orthogonally oriented dove prisms (DPs). Cameras (ii) and (iii) detect images identical in shape but half as bright as those captured by cameras (iii,iv) in the dual-objective system in \Cref{fig:dualObjSys}. (b,c) Images of molecules with orientations $-\mu_x=\mu_y=\mu_z$ and $\mu_x=\mu_y=\mu_z$ captured at detectors (i)-(iii) and intermediate image planes (iv,v) (b)~without and (c)~with VWPs. Images depict a total of 2000 photons detected. Scale bar: 1~$\mathrm{\mu m}$. Colorbar: photons per $58.5 \times 58.5~\text{nm}^2$ pixel.}
    \label{fig:singleObj_FI_basis}
\end{figure}

Here, we show a single-objective interferometric imaging system that achieves QCRB-limited precision for estimating first-order orientational dipole moments (\Cref{fig:singleObj_FI_basis}), analogous to the dual-objective system discussed in the main text (\Cref{fig:dualObjSys}). This system similarly uses a vortex waveplate (VWP) to circumvent the need for interferometric detection of azimuthally polarized emission light. However, this system passes radially polarized light through two 50/50 beamsplitters in a Mach-Zehnder configuration. Each arm further uses a dove prism (DP) to flip the field for proper detection of orientation information. 

Although this imaging system is simpler to implement than a dual-objective system, the use of only one objective lens prevents cameras (iii) and (iv) from measuring the position $(x,y)$ and orientation $(\mu_x,\mu_y)$ simultaneously (\Cref{fig:singleObj_FI_basis}). For single-objective detection, the $y$-polarized field at the BFP for a molecule located at position $(x,y)$ is given by
\begin{equation}
\label{eqn:bfp_pos}
    \psi'_{y}(u,v;x,y,\bm{\mu})=\psi_y(u,v;\bm{\mu})\exp[jk(ux+vy)],
\end{equation}
whereas for the dual-objective system, the electric fields collected by objectives 1 and 2 are given by
\begin{linenomath} \begin{subequations}
\begin{align}
    \psi'_{y,1}(u_1,v_1;x,y,\bm{\mu})&=\psi_y(u_1,v_1;\bm{\mu})\exp[jk(u_1x+v_1y)]\\
    \psi'_{y,2}(u_2,v_2;x,y,\bm{\mu})&=\psi_y(u_2,v_2;\bm{\mu})\exp[jk(-u_2x-v_2y)].
\end{align}
\end{subequations} \end{linenomath} 

As stated in the main text (\Cref{sec:reachingQCRB}), orientation measurements in the image plane achieve maximum precision when even and odd fields at the BFP are separated, e.g., when $\psi_y(u,v)+\psi_y(-u,-v)$ and $\psi_y(u,v)-\psi_y(-u,-v)$ are resolved simultaneously. In the dual-objective setup in \Cref{fig:dualObjSys}, the fields captured by cameras (iii) and (iv) are given by $\psi'_{y,1}(u,v)+\psi'_{y,2}(-u,-v)$ and $\psi'_{y,1}(u,v)-\psi'_{y,2}(-u,-v)$, respectively. Thereby, the orientation measurement is optimized, and position information $\exp[jk(ux+vy)]$ is preserved. 

However, for single-objective detection, we can only optimize the orientation measurement for a single $(x,y)$ position, e.g., for $(x,y)=(0,0)$ by interfering $\psi_y'(u,v;0,0,\bm{\mu})$ with $\psi_y'(-u,-v;0,0,\bm{\mu})$ using DPs as depicted in \Cref{fig:singleObj_FI_basis}. Therefore, position information $\exp[jk(ux+vy)]$ is lost, but an image may be formed point-by-point by scanning the illumination or sample over time.

\section{Quantum Fisher information of estimating second-order orientational dipole moments}
\label{appendix:secondOrder}
For a non-fixed, i.e., rotationally diffusing, molecule, the photon density matrix is given by
\begin{linenomath} \begin{multline}
    \rho = \op{g_x}M_{xx} +\op{g_y}M_{yy} + (\op{g_x}{g_y}+\op{g_y}{g_x})M_{xy} \\
    + (\op{g_x}{g_z}+\op{g_z}{g_x})M_{xz} + (\op{g_y}{g_z}+\op{g_z}{g_y})M_{yz} \\
    + \big[\op{g_z}+(1-c)\op{\text{vac}}\big]M_{zz},
\end{multline} \end{linenomath} 
where $\ket{g_i}$ are the basis fields measured in the BFP [\Cref{eqn:basisFieldsBFP}] and $M_{ij}=(1/T)\int_0^T\mu_i\mu_j\,dt=\langle \mu_i\mu_j \rangle$ is the temporal average of the second moments of molecular orientation over acquisition time $T$. The partial derivatives of the density matrix with respect to the orientational second-order moments are written as
\begin{linenomath} \begin{subequations}
\begin{align}
    \pdv{\rho}{M_{ii}}&=\op{g_i} &i\in\{x,y,z\}\\
    \pdv{\rho}{M_{ij}}&=\op{g_i}{g_j}+\op{g_j}{g_i} &i,j\in\{x,y,z\},\,i\neq j.
\end{align}
\end{subequations} \end{linenomath}
Note that while we parameterize the molecule's orientation and rotational diffusion using six second-order orientational moments $M_{ij}$, generalized Stokes parameters $S_i$ \cite{brosseau1998fundamentals,curcio2019birefringent} and spherical harmonics \cite{Chandler:19ii} may be used instead via a change of variables. For example, the generalized Stokes parameters $S_i$ may be computed in terms of the second moments as follows:
\begin{linenomath} \begin{subequations}
\begin{align}
    &S_1 = 2S_0(M_{xx}-M_{yy})\\
    &[S_2,S_4,S_6]=2S_0[M_{xy},M_{xz},M_{yz}]\\
    &S_3=S_5=S_7=0\\
    &S_8 = \sqrt{3}S_0(M_{xx}+M_{yy}-M_{zz})/2,
\end{align}
\end{subequations} \end{linenomath}
where brightness scaling factor $S_0=1$ for 1 photon detected. A linear transformation can be applied to project our results into this space.

The SLDs and QFI can be computed numerically for any orientation $[M_{xx},M_{yy},M_{zz},M_{xy},M_{xz},M_{yz}]$; due to the complexity of the eigendecomposition of $\rho$, it is difficult to find a simple analytical expression. However, for a molecule symmetrically wobbling around the $\mu_z$-axis with rotational constraint $\gamma$, we may write the density matrix as
\begin{linenomath} \begin{multline}
    \rho =\frac{1-\gamma}{3}(\op{g_x}+\op{g_y}) \\
    +\frac{1+2\gamma}{3} \big[\op{g_z}+(1-c)\op{\text{vac}}\big].
\end{multline} \end{linenomath}
The SLDs corresponding to the second-order moments become
\begin{linenomath} \begin{subequations}
\begin{align}
    \mathcal{L}_{{ii}}&=\frac{3}{1+2\gamma}\op{g_i}, \qquad \qquad i\in{\{x,y\}} \label{eqn:SLD_ii} \\
    \mathcal{L}_{{zz}}&=\frac{3}{c(1+2\gamma)}\op{g_z} \label{eqn:SLD_zz}\\
    \mathcal{L}_{{xy}}&=\frac{3}{1-\gamma}(\op{g_x}{g_y}+\op{g_y}{g_x}) \label{eqn:SLD_xy}\\
    \mathcal{L}_{{iz}}&=\frac{6}{1+c+(2c-1)\gamma}(\op{g_i}{g_z}+\op{g_z}{g_i}) \label{eqn:SLD_iz}.
\end{align} 
\end{subequations} \end{linenomath}
The QFI can be computed according to \Cref{eqn:QFI}, yielding \Cref{eqn:QFI2nd_z} and
\begin{equation}
    \mathcal{K}^z_{xz,xz}=\mathcal{K}^z_{yz,yz}=\frac{12c}{1+c+(2c-1)\gamma}.
    \label{eqn:QFI2nd_z_yz}
\end{equation}
A similar procedure can be applied to compute the QFI of $x$-oriented molecules, yielding \Cref{eqn:QFI2nd_x} and 
\begin{linenomath} \begin{subequations}
\begin{align}
    \mathcal{K}^x_{xz,xz}&=\frac{12c}{1+c+(2-c)\gamma}\\
    \mathcal{K}^x_{yz,yz}&=\frac{12c}{(1+c)(1-\gamma)}.
\end{align}
\label{eqn:QFI2nd_x_yz}
\end{subequations} \end{linenomath}

Note that a measurement that projects onto the SLDs corresponding to the squared moments (\Cref{eqn:SLD_ii,eqn:SLD_zz}), which is sufficient to achieve QCRB-limited precision for measuring those moments, requires $\ket{g_x}$, $\ket{g_y}$ and $\ket{g_z}$ to be resolved separately on a camera. In contrast, a measurement that projects onto the SLDs corresponding to the cross moments (\Cref{eqn:SLD_xy,eqn:SLD_iz}) requires $\ket{g_x}$, $\ket{g_y}$ and $\ket{g_z}$ to overlap with one another on the camera. 

Interestingly, although $\sigma_{xx,\text{QCRB}}$, $\sigma_{yy,\text{QCRB}}$, and $\sigma_{xy,\text{QCRB}}$ vary with the mean azimuthal orientation $\bar{\phi}$ [\Cref{fig:QCRB_CRB_2nd_fixed}(b)], the SGV $\sigma_{xx,yy,\text{QCRB}}^2$ representing the area of the covariance ellipse for estimating [$M_{xx},M_{yy}]$ is constant and uniform for all $\bar{\phi}$ [\Cref{fig:QCRB_CRB_2nd_fixed}(a)]. 

\section{Classical Fisher information of estimating second-order orientational dipole moments}
\label{appendix:2ndOrderClassical}

\begin{figure}[htbp]
    \centering
    \includegraphics{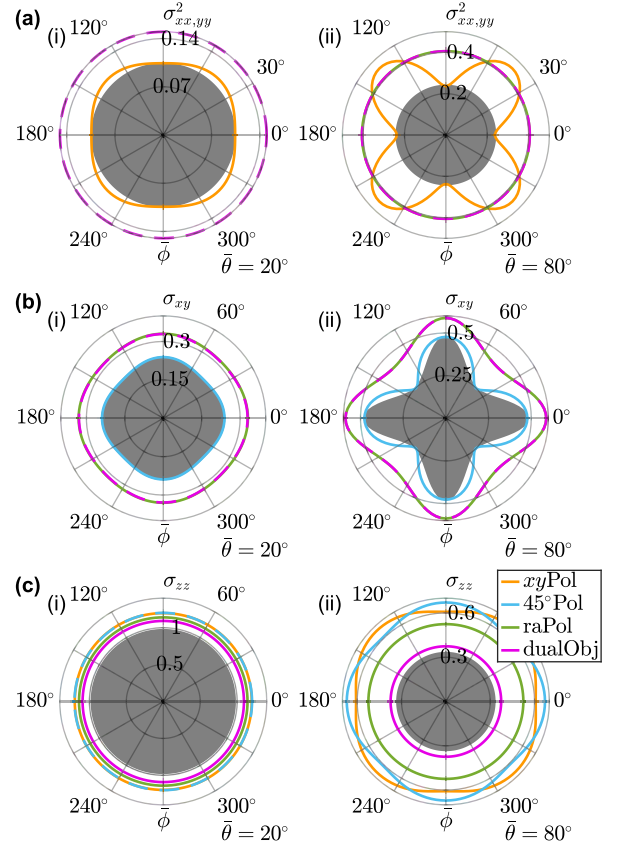}
    \caption{Classical CRB of several techniques (\Cref{appendix:2ndOrderClassical}) compared to the quantum CRB of estimating second-order orientational moments of a nearly fixed ($\gamma=0.8$) dipole emitter with average polar orientation (i)~$\bar{\theta}=20^\circ$ and (ii)~$\bar{\theta}=80^\circ$ using 1 detected photon. (a) CRB SGV for estimating in-plane squared moments $M_{xx}$ and $M_{yy}$ using a low 0.1 NA objective. (b) Best-possible precision $\sqrt{\text{CRB}}$ of estimating the in-plane cross moment $M_{xy}$ using a low 0.1 NA objective. (c)~Best-possible precision $\sqrt{\text{CRB}}$ of estimating the out-of-plane squared moment $M_{zz}$ using a high 1.4 NA objective. The gray regions are bounded from above by the numerically computed (a)~QCRB or (b,c)~$\sqrt{\text{QCRB}}$. Orange: standard PSF with $x$- and $y$-polarized detection ($xy$Pol), cyan: standard PSF with linearly polarized detection at $\pm 45^\circ$ in the $xy$-plane ($45^\circ$Pol), green: standard PSF with radially  and azimuthally polarized detection (raPol), and magenta: dual-objective interferometric detection with VWPs (dualObj). All curves assume 1 photon is detected from the dipole emitter. The estimation precision of $45^\circ$Pol in (a) and $xy$Pol in (b) are orders of magnitude larger than those of the other techniques and are not shown.}
    \label{fig:QCRB_CRB_2nd_fixed}
\end{figure}

We expand the classical image formation model $I(u,v)=\Psi\Psi^*$ in terms of the second moments of molecular orientation as
\begin{linenomath} \begin{multline}
    I(u,v) = s\big[ B_{xx}(u,v) M_{xx} + B_{yy}(u,v) M_{yy} \\
    + B_{zz}(u,v) M_{zz} + B_{xy}(u,v) M_{xy} \\
    + B_{xz}(u,v) M_{xz} + B_{yz}(u,v) M_{yz} \big],
    \label{eqn:classical_image_formation_2nd}
\end{multline} \end{linenomath} 
where $B_{ij}(u,v)$ are the intensity basis images given by
\begin{linenomath} \begin{subequations}
\label{eqn:basis_image}
\begin{align}
    B_{ii}&=G_iG_i^*&i\in\{x,y,z\}\\
    B_{ij}&=G_iG_j^*+G_i^*G_j&i,j\in\{x,y,z\},\,i\neq j,
\end{align}
\label{eqn:basis_images_classical}
\end{subequations} \end{linenomath}
$G_i$ are the basis fields in the image plane [\Cref{eqn:basisFieldsImg}], and $s=1$ is a brightness scaling factor corresponding to one photon detected. 

To investigate the trade-off in measuring squared vs. cross moments, we analyze the in-plane second order moments $(M_{xx},M_{yy},M_{xy})$ and assume that $B_{xz}=B_{yz}=B_{zz}=0$ for simplicity. Since the total intensity of an image must be non-negative everywhere, the inequality
\begin{equation}
    I=B_{xx}M_{xx}+B_{yy}M_{yy}+B_{xy}M_{xy}\geq0,
\end{equation}
must be satisifed for all $(M_{xx},M_{yy},M_{xy})$ such that $M_{xy}^2 = \langle \mu_x \mu_y \rangle^2 \leq \langle \mu_x^2 \rangle \langle \mu_y^2 \rangle = M_{xx}M_{yy}$. From the definition of the intensity basis images [\Cref{eqn:basis_image}], we have
\begin{linenomath} \begin{align}
    B_{xy}^2&=(G_xG_y^*+G_x^*G_y)^2 \nonumber\\
    &=\,2G_xG_x^*G_yG_y^*+(G_xG_y^*)^2+(G_x^*G_y)^2 \nonumber\\
    &=\,2G_xG_x^*G_yG_y^*+2\Re{ (G_xG_y^*)^2 } \nonumber\\
    \leq\,&2G_xG_x^*G_yG_y^*+2\abs{G_xG_y^*}^2 \nonumber\\
    &=\,4G_xG_x^*G_yG_y^*=4B_{xx}B_{yy}, \label{eqn:BxyBoundBxxByy}
\end{align} \end{linenomath}
with equality if and only if $G_xG_y^*$ is real, i.e., $G_x$ and $G_y$ have the same phase. Note that this inequality holds for all imaging systems, i.e., any possible $G_i$.

The classical FI matrix of estimating in-plane orientational second moments, ignoring the third, fifth, and sixth rows and columns of the full FI matrix $\bm{\mathcal{J}}$, may be written as
\begin{linenomath} \begin{align}
    \bm{\mathcal{J}}_{xx,yy,xy} &= \begin{bmatrix} \mathcal{J}_{xx,xx} & \mathcal{J}_{xx,yy} & \mathcal{J}_{xx,xy} \\
    \mathcal{J}_{yy,xx} & \mathcal{J}_{yy,yy} & \mathcal{J}_{yy,xy} \\
    \mathcal{J}_{xy,xx} & \mathcal{J}_{xy,yy} & \mathcal{J}_{xy,xy} \end{bmatrix} \nonumber \\
    &= \begin{bmatrix} \mathcal{J}_{11} & \mathcal{J}_{12} & \mathcal{J}_{14} \\
    \mathcal{J}_{21} & \mathcal{J}_{22} & \mathcal{J}_{24} \\
    \mathcal{J}_{41} & \mathcal{J}_{42}& \mathcal{J}_{44} \end{bmatrix}.
\end{align} \end{linenomath}
We use the square root of the inverse of the determinant of the $2\times2$ FI submatrix $\bm{\mathcal{J}}_{xx,yy}$ to quantify the CRB SGV of estimating the squared second moments [\Cref{fig:QCRB_CRB_2nd_NA_gamma}(a)], and we invert the diagonal entry $\mathcal{J}_{44}$ to compute the CRB corresponding to $M_{xy}$ [\Cref{fig:QCRB_CRB_2nd_NA_gamma}(b)]. When using the aforementioned polarized standard PSFs ($xy$Pol and raPol, with and without interferometric detection) to measure molecules with mean orientations $\bar{\mu}_x=1$ or $\bar{\mu}_z=1$, there is zero correlation between estimating in-plane squared moments ($M_{xx}$, $M_{yy}$) and the cross moment ($M_{xy}$), i.e., $\mathcal{J}_{14}=\mathcal{J}_{24}=0$; thus the diagonal entry $\mathcal{J}_{44}$ can be directly evaluated for quantifying classical FI.

Next, we compute the classical FI of measuring the in-plane second moments of a molecule wobbling around the $\mu_x$ axis, i.e., $M_{xx}=(1+2\gamma)/3$, $M_{yy}=M_{zz}=(1-\gamma)/3$ and $M_{xy}=0$, as
\begin{linenomath} \begin{subequations}
\begin{align}
\label{eqn:classicalMxx}
    \mathcal{J}_{11}&=\iint\frac{3B_{xx}^2}{(1+2\gamma)B_{xx}+(1-\gamma)B_{yy}}\,du\,dv\nonumber\\*
    &\leq\iint \frac{3B_{xx}}{1+2\gamma}\,du\,dv=\frac{3}{1+2\gamma}=\mathcal{K}_{xx,xx}\\
\label{eqn:classicalMyy}
    \mathcal{J}_{22}&=\iint\frac{3B_{yy}^2}{(1+2\gamma)B_{xx}+(1-\gamma)B_{yy}}\,du\,dv\nonumber\\*
    &\leq\iint \frac{3B_{yy}}{1-\gamma}\,du\,dv=\frac{3}{1-\gamma}=\mathcal{K}_{yy,yy}\\
\label{eqn:classicalMxxyy}
    \mathcal{J}_{12}&=\iint\frac{3B_{xx}B_{yy}}{(1+2\gamma)B_{xx}+(1-\gamma)B_{yy}}\,du\,dv\\
\label{eqn:classicalMxy}
    \mathcal{J}_{44}&=\iint\frac{3B_{xy}^2}{(1+2\gamma)B_{xx}+(1-\gamma)B_{yy}}\,du\,dv\nonumber\\*
    &\leq\iint\frac{12B_{xx}B_{yy}}{(1+2\gamma)B_{xx}+(1-\gamma)B_{yy}}\,du\,dv=4\mathcal{J}_{12}.
\end{align}
\end{subequations} \end{linenomath}
We now develop a relation between the covariance $\mathcal{J}_{12}$ and the diagonal elements $\mathcal{J}_{11}$ and $\mathcal{J}_{22}$ given by \Cref{eqn:classicalMxx,eqn:classicalMyy,eqn:classicalMxxyy}, yielding
\begin{equation}
    \frac{1+2\gamma}{1-\gamma}(\mathcal{K}_{xx,xx}-\mathcal{J}_{11})=\frac{1-\gamma}{1+2\gamma}(\mathcal{K}_{yy,yy}-\mathcal{J}_{22})=\mathcal{J}_{12}
\end{equation}
where we have utilized the fact that the total energies in $B_{xx}$ and $B_{yy}$ are each normalized to one. The equalities in \Cref{eqn:classicalMxx,eqn:classicalMyy} are only satisfied when $B_{xx}B_{yy}=0\ \forall\ [u,v]$, i.e., the classical FI saturates the QFI when $B_{xx}$ and $B_{yy}$ are spatially separated on the camera. However, if this condition holds, then $B_{xy}=0$ [\Cref{eqn:BxyBoundBxxByy}], i.e., $I$ does not depend on $M_{xy}$, and $I$ does not contain any information for measuring $M_{xy}$.

In the main text, we discussed a trade-off between achieving good precision in estimating squared second moments, e.g., $M_{xx}$, versus achieving good precision in estimating cross-moments, e.g., $M_{xy}$, for molecules wobbling around the in-plane axes or the optical axis. Here, we compute numerically the precision of measuring second moments for molecules with arbitrary average orientations $(\bar{\theta},\bar{\phi})$ and small rotational diffusion ($\gamma=0.8$), which is equivalent to rotating uniformly within a cone of half-angle $30.7^\circ$, using various methods [\Cref{fig:QCRB_CRB_2nd_fixed}]. The estimation precisions for mostly fixed molecules are similar to those for freely rotating molecules (\Cref{fig:QCRB_CRB_2nd_NA_gamma}). The $x/y$-polarized standard PSF with a low NA objective lens has a precision achieving the quantum bound for measuring $M_{xx}$ and $M_{yy}$ for some orientations, but has no sensitivity for measuring $M_{xy}$. The $45^\circ$-polarized standard PSF has the opposite performance; it achieves the QCRB for measuring $M_{xy}$, but has no sensitivity for measuring $M_{xx}$ and $M_{yy}$. The radially/azimuthally polarized standard PSF has better $M_{zz}$ precision compared to the in-plane polarized PSFs. We surmise that these methods do not simultaneously achieve QCRB-limited precision for all orientations because they do not project onto the corresponding SLDs for the orientational second moments.

We next consider the CRB SGV $(\sigma^x_{xx,yy,xy})^2$ for estimating $M_{xx}$, $M_{yy}$ and $M_{xy}$ simultaneously, given by
\begin{linenomath} \begin{multline}
    \left(\sigma^x_{xx,yy,xy}\right)^2 = [\det(\bm{\mathcal{J}}_{xx,yy,xy})]^{-1/3} \\
    =[\det(\bm{\mathcal{J}}_{xx,yy})\mathcal{J}_{44} -\mathcal{J}_{11}\mathcal{J}_{24}^2 \\*
    -\mathcal{J}_{22}\mathcal{J}_{14}^2+2\mathcal{J}_{12}\mathcal{J}_{14}\mathcal{J}_{24}]^{-1/3}.
\end{multline} \end{linenomath}
We derive a bound for the off-diagonal FI elements as
\begin{linenomath} \begin{multline}
    \mathcal{J}_{11}\mathcal{J}_{24}^2+\mathcal{J}_{22}\mathcal{J}_{14}^2-2\mathcal{J}_{12}\mathcal{J}_{14}\mathcal{J}_{24} \\ 
    \geq \, 2\sqrt{\mathcal{J}_{11}\mathcal{J}_{22}} \abs{\mathcal{J}_{14}\mathcal{J}_{24}} - 2\abs{\mathcal{J}_{12}\mathcal{J}_{14}\mathcal{J}_{24}} \\
    = \, 2\left(\sqrt{\mathcal{J}_{11}\mathcal{J}_{22}}-\abs{\mathcal{J}_{12}}\right)\abs{\mathcal{J}_{14}\mathcal{J}_{24}}\geq0,
\end{multline} \end{linenomath}
assuming that the FI submatix for estimating $M_{xx}$ and $M_{yy}$ is positive definite, i.e., $\mathcal{J}_{11}\mathcal{J}_{22}-\mathcal{J}_{12}>0$, since the SGV becomes infinite if any determinant of any of the FI submatrices is 0. Equality holds if and only if $\mathcal{J}_{14}=\mathcal{J}_{24}=0$, i.e., measurements of $M_{xx}$ and $M_{yy}$ are uncorrelated with $M_{xy}$. Therefore, the SGV is bounded as
\begin{linenomath} \begin{align}
\label{eqn:classicalBound}
    \left(\sigma^x_{xx,yy,xy}\right)^2 &= [\det(\bm{\mathcal{J}}_{xx,yy,xy})]^{-1/3} \nonumber\\
    &\geq [\det(\bm{\mathcal{J}}_{xx,yy})\mathcal{J}_{44}]^{-1/3} \nonumber\\
    &\geq \left[4(\mathcal{J}_{11}\mathcal{J}_{22}-\mathcal{J}_{12}^2)\mathcal{J}_{12}\right]^{-1/3} \nonumber\\
    &\geq \frac{[(1-\gamma)(2+\gamma)(2\gamma+1)]^{1/3}}{3}\nonumber\\
    &= 4^\frac{1}{3}\left(\sigma_{xx,yy,xy,\text{QCRB}}^x\right)^2,
\end{align} \end{linenomath}
where the minimum SGV in the final inequality is found by setting $\pdv*{(\sigma^x_{xx,yy,xy})^2}{\mathcal{J}_{11}}=0$. Similarly, for $z$-oriented molecules, the SGV is bounded by
\begin{equation}
\label{eqn:classicalBound_z}
    \left(\sigma_{xx,yy,xy}^z\right)^2\geq 2^{\frac{1}{3}}\,\frac{1-\gamma}{3}=4^\frac{1}{3}\left(\sigma_{xx,yy,xy,\text{QCRB}}^z\right)^2.
\end{equation}

\begin{figure}
    \centering
    \includegraphics{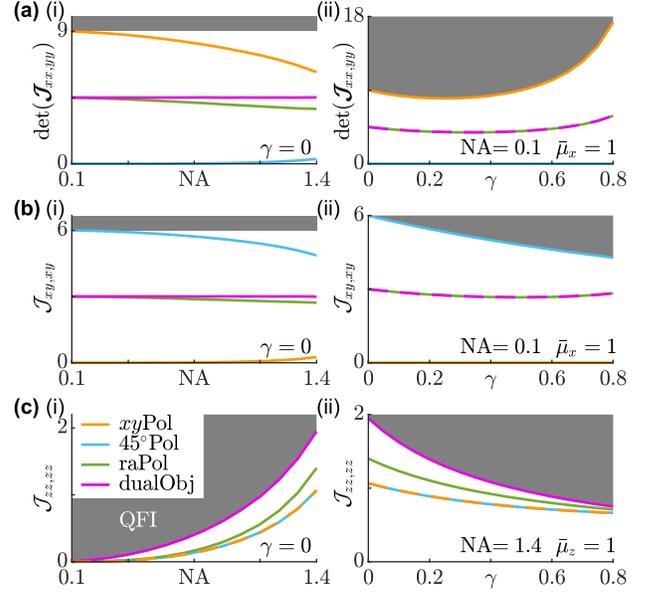}
    \caption{Classical FI of several techniques (\Cref{appendix:2ndOrderClassical}) versus quantum FI of estimating second-order orientational moments of dipole emitters. (a) Inverse of generalized variance for estimating $M_{xx}$ and $M_{yy}$, classical FIs for estimating (b) $M_{xy}$ for molecules wobbling around the $\mu_x$ axis and (c) $M_{zz}$ for molecules wobbling around the $\mu_z$ axis as functions of (i)~numerical aperture NA (for $\gamma=0$) and (ii)~rotational constraint $\gamma$ [for (a),(b) $\text{NA}=0.1$ and (c) $\text{NA}=1.4$]. The gray regions are bounded from below by the QFI [\Cref{eqn:QFI2nd}]. Orange: standard PSF with $x$- and $y$-polarized detection ($xy$Pol), cyan: standard PSF with linearly polarized detection at $\pm 45^\circ$ in the $xy$-plane ($45^\circ$Pol), green: standard PSF with radially  and azimuthally polarized detection (raPol), and magenta: dual-objective interferometric detection with VWPs (dualObj). All curves assume 1 photon is detected from the dipole emitter.}
    \label{fig:QFI_FI_2nd_NA_gamma}
\end{figure}

We therefore observe that the classical CRB for measuring in-plane second moments is bounded; the precision in measuring $M_{xx}$, $M_{yy}$, and $M_{xy}$ cannot simultaneously reach the best-possible QCRB. These tradeoffs are exemplified by comparing the $xy$Pol and $45^\circ$Pol techniques in \Cref{fig:QCRB_CRB_2nd_NA_gamma}(a,b), \Cref{fig:QCRB_CRB_2nd_fixed}(a,b), and \Cref{fig:QFI_FI_2nd_NA_gamma}(a,b). Interestingly, although both of these methods saturate the QCRB for subsets of $M_{xx}$, $M_{yy}$, and $M_{xy}$, their SGV for measuring all in-plane moments is poor. In contrast, raPol and dual-objective techniques cannot saturate the QCRB for any one in-plane second moment, but their SGV for all in-plane moments is very close to the bound given by \Cref{eqn:classicalBound,eqn:classicalBound_z} [\Cref{fig:QCRB_CRB_2nd_NA_gamma}(a,b), \Cref{fig:classicalCRB_tradeoff}, and \Cref{fig:QFI_FI_2nd_NA_gamma}(a,b)]. This analysis can be extended to $z$-related squared and cross moments, resulting in a similar trade-off.

\section{Impact of background photons on the estimation precision of first-order orientational dipole moments}
\label{appendix:background}
In this section, we briefly discuss the effect of background on the estimation precision. The estimation precision in the presence of background highly depends on the nature of the background photons, especially their spatial distributions. We write a new density matrix $\rho$, accounting for signal and background emitters, as
\begin{linenomath} \begin{multline}
\label{eqn:density_bg}
    \rho =s\left[(1-\epsilon_z)\op{\text{vac}}+\op{\psi}\right]+b_\perp\rho_\perp \\*  
    +\frac{b}{3}(\op{g_x}+\op{g_y}+c^{-1}\op{g_z}),
\end{multline} \end{linenomath} 
where $s$ represents the fraction of photons from the dipole of interest, $b_\perp$ represents the fraction of photons from background sources that are orthogonal to the basis fields $\ket{g_i}$ with density matrix $\rho_\perp$, and $b$ represents the background photons that project uniformly onto the basis fields $\ket{g_i}$. Here, we assume background sources $b_\perp$ do not contaminate the orientation measurement, while background sources $b$ will affect the measurement. The summed contributions of signal and background photons must be normalized, i.e., $s+b_\perp+b=1$.

\begin{figure}[htbp]
    \centering
    \includegraphics{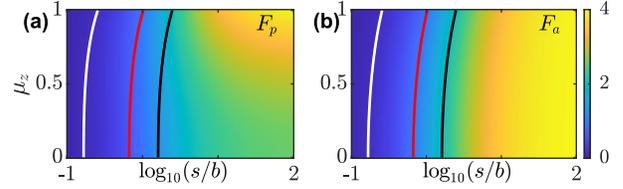}
    \caption{Quantum FI of estimating first-order orientational moments of fixed dipole emitters as a function of signal-to-background ratio (SBR). (a) QFI of estimating polar orientation. (b) QFI of estimating azimuthal orientation. Black, red, and white lines represent a QFI reduction of 50, 75, and 96 percent, i.e., a best-possible standard deviation in the presence of background equal to $\sqrt{2}$ times, twice, and 5 times that without background, respectively.}
    \label{fig:fig6}
\end{figure}

Similar to the backgroundless case [\Cref{eqn:QFI_1st}], the QFI is also azimuthally symmetric, given by
\begin{equation}
    \bm{\mathcal{K}}=s\left(\frac{F_p}{\mu_z^2}\bm{\nu}_p\bm{\nu}_p^\dagger+F_a\bm{\nu}_a\bm{\nu}_a^\dagger\right),
\end{equation}
where
\begin{linenomath} \begin{subequations}
\begin{align}
    F_p &= \frac{4}{\epsilon_z}\left\{\frac{(\epsilon_z-c)(1-\epsilon_z)}{\left[1+b/(3s\epsilon_z)\right]^2}+\frac{c}{\left[1+2b/(3s\epsilon_z)\right]^2}\right\} \\
    F_a &=4\left(1+\frac{2b}{3s\epsilon_z}\right)^{-2},
\end{align}
\end{subequations} \end{linenomath} 
and the probability of a photon emitted by the dipole that escapes detection is given by $1-\epsilon_z=(1-c)\mu_z^2$. Compared to the backgroundless case and averaging over all possible orientations, the best-possible precision decreases by a factor of two for a signal-to-background ratio (SBR) $s/b=0.75$ (\Cref{fig:fig6}), i.e., 3 background photons are detected for every 4 signal photons. Note that we have assumed that these background photons project uniformly across $\ket{g_x}$, $\ket{g_y}$, and $\ket{g_z}$ in \Cref{eqn:density_bg}. The QCRB will change depending upon how photons from the background emitters project onto the basis fields of the imaging system.


\bibliography{references}

\end{document}